\documentclass{article}

\usepackage{graphicx,
            amsmath,
            amsthm,
            hyperref,
            amssymb,
            enumerate,
            tikz}

\usepackage[linesnumbered,ruled,vlined]{algorithm2e}
\usepackage[margin=1.5in]{geometry}
\usepackage{xcolor}
\usepackage{float}
\allowdisplaybreaks

\usepackage{libertine}
\usepackage[libertine]{newtxmath}
\usetikzlibrary{positioning}

\usepackage{mathtools}

\SetKw{Continue}{continue}
\SetKw{Break}{break}

\newtheorem{theorem}{Theorem}
\newtheorem{definition}{Definition}
\newtheorem{lemma}{Lemma}
\newtheorem{proposition}{Proposition}

\newcommand{\ds}{\textup{d}}
\newcommand{\Opt}{\text{OPT}}

\begin{document}

\title{Fast and Simple Densest Subgraph with Predictions}

\author{
Thai Bui\thanks{tbui8182@sdsu.edu, San Diego State University. Supported by NSF Grant No. 2342527.},  
Luan Nguyen \thanks{lnguyen9027@sdsu.edu, San Diego State University.}, \\
Hoa T. Vu\thanks{hvu2@sdsu.edu, San Diego State University. Supported by NSF Grant No. 2342527.},
}

  \maketitle

\begin{abstract}

We study the densest subgraph problem and its NP-hard densest at-most-$k$ subgraph variant through the lens of learning-augmented algorithms. We show that, given a reasonably accurate predictor that estimates whether a node belongs to the solution (e.g., a machine learning classifier), one can design simple linear-time algorithms that achieve a $(1-\epsilon)$approximation. Finally, we present experimental results demonstrating the effectiveness of our methods for the densest at-most-$k$ subgraph problem on real-world graphs.

\end{abstract}

\section{Introduction}
Computing the densest subgraph and its variants is a well-studied problem in graph algorithms and data mining. Given a graph $G=(V,E)$, the density of a subgraph induced by a subset of nodes $H$ is defined as 
\[
    \ds(H)=\frac{|E(H)|}{|H|}.
\] 

The densest subgraph problem asks to find the subgraph with the maximum possible density. Other interesting variants of this problem include the densest $k$ subgraph, densest at-most-$k$ subgraph, and densest at-least-$k$ subgraph problems where $H$ is required to have size $k$, $\le k$, $\ge k$ respectively \cite{Manurangsi17, BhaskaraCCFV10, KhullerS09, AndersenC09}.

The densest subgraph problem, initially explored by Picard and Queyranne \cite{PQ82} and Goldberg \cite{Goldberg84}, has since become central to a wide range of applications, including community detection in social networks \cite{DGP07, CS12}, link spam detection \cite{GKT05, BXGPF13}, and computational biology \cite{FNBB06, SHKRZ10}. Over time, faster algorithms have been proposed for the problem and its variants \cite{GGT89, Charikar00, KS09}. Additionally, the problem has been extensively studied across various computational models, including streaming \cite{BHNT15, EsfandiariHW15, McGregorTVV15}, dynamic \cite{SJ20, BHNT15}, massively parallel \cite{BahmaniKV12, BahmaniGM14, GLM19}, and distributed settings \cite{SarmaLNT12, SuV20}. For a survey, see \cite{LancianoMFB24}.

In this work, we study the densest subgraph problem and its NP-hard variant, the densest at-most-$k$ subgraph, within the framework of learning-augmented algorithms. This approach is motivated by the observation that real-world data often exhibit underlying patterns that can be captured using machine learning. We mainly assume access to a predictor (e.g., a machine-learning model) that estimates whether a node belongs to the densest subgraph. Such a predictor is typically obtained by training a model on a dataset for which ground-truth labels are available.

Designing algorithms that incorporate machine learning predictions has seen a lot of recent success and became a popular paradigm. Examples of graph problems that have been studied in this paradigm are shortest paths, maximum matching, maximum independent set, min cut, and triangle counting \cite{FeijenGuido2021,BravermanDSW24, ChenSVZ22, MNS2025, Dong0V25, BoldrinV24}. Many other problems in algorithms and data structures in different computational models have also been studied in this setting. For a survey, see \cite{MitzenmacherV22}. A more comprehensive list of works can be found at \cite{algorithms_with_predictions}.

In the context of the densest subgraph problem, a predictor can be viewed as a classifier that predicts whether a node belongs to the densest subgraph, based on features such as its degree, average neighbor degree, and domain-specific attributes, if available.

\paragraph{The pitfall of outputting the predictor's solution.} However, blindly relying on the predictor's output might result in a poor solution. The predictor may simply have low accuracy in predicting which nodes belong to the densest subgraph. This issue can often be overcome in practice by rigorous model testing and training.

More dangerously, even an accurate predictor can still lead to poor approximations if it is not combined with additional algorithmic processing. One can devise a simple example in the context of the densest subgraph problem. Suppose that we have a predictor that correctly identifies 90\% of the nodes in the densest subgraph. If the densest subgraph is a dense bipartite graph with partitions $X,Y$ and $|Y| = 0.9 (|X| + |Y|)$ (i.e., $Y$ contains 90\% of the nodes of the densest subgraph), and the predictor provides us with $Y$, simply returning $Y$ results in a subgraph with density 0, which is arbitrarily worse than the optimal solution, even though the predictor has a 90\% ``recall''. See Figure \ref{fig:1} for an example.

\begin{center}
\begin{figure}
    \centering

    \begin{tikzpicture}[node distance=0.8cm]
    
      \node[circle, fill=black, inner sep=1.5pt, label=left:$x_1$] (x1) at (0,2) {};
      \node[circle, fill=black, inner sep=1.5pt, label=left:$x_2$] (x2) at (0,0) {};
    
      \foreach \i in {1,...,5} {
        \node[circle, fill=black, inner sep=1.5pt, label=right:$y_{\i}$] 
          (y\i) at (3,4.5-1*\i) {};
      }
    
      \foreach \i in {1,2} {
        \foreach \j in {1,...,5} {
          \draw (x\i) -- (y\j);
        }
      }
    
    \end{tikzpicture}
    \caption{Suppose that this is the densest subgraph and the predictor identifies the nodes $y_1,\ldots,y_5$. Returning $y_1,\ldots,y_5$ will yield an infinitely bad solution with density 0.}
    \label{fig:1}
\end{figure}

\end{center}

\paragraph{Related work.} Exact algorithms for the densest subgraph problem exist using linear programming or maximum flow \cite{Charikar00, Goldberg84}. For unweighted graphs, Goldberg's max-flow based algorithm runs in $O(mn \log n)$ time \cite{LancianoMFB24}. One can also approximate the densest subgraph up to a $1-\epsilon$ factor using the multiplicative weights method to solve the dual linear program followed by a clever rounding algorithm \cite{SuV20, BahmaniGM14, GhaffariLM19} which runs in $O(m\cdot 1/\epsilon^2 \cdot \log^2 n)$ time. There is, however, a simple and practical greedy $1/2$-approximation that can be implemented in $O(m+n)$ time by Charikar \cite{Charikar00}; this algorithm keeps removing the node with the lowest degree from the graph and recording the densest subgraph in the process. 

There is also a heuristic proposed by Boob et al. \cite{BoobGPSTWW20} called Greedy++; this heuristic was later shown by Chekuri, Quanrud, and Torres \cite{ChekuriQT22} to run in $O\left( \frac{m \Delta \log n}{\epsilon^2 d(H^\star)} \right)$ time (which is $O(\epsilon^{-2}mn\log n)$ in the worst case).  Chekuri, Quanrud, and Torres \cite{ChekuriQT22} also provided a flow-based $\tilde{O}(m/\epsilon)$ time algorithm that obtains a $1-\epsilon$ approximation.

One {\em practical issue} with the densest subgraph problem is that the densest subgraph is often too large to be useful in many real-world graphs. Instead, one may consider the problem of finding the densest subgraph with at most $k$ nodes. This variant is NP-hard. Andersen \cite{Andersen} showed that an $\alpha$-approximation for this problem implies an $O(\alpha^2)$-approximation for the densest $k$ subgraph problem. Furthermore, under the Exponential Time Hypothesis (ETH), no polynomial-time constant-factor approximation exists for the densest $k$ subgraph problem \cite{Manurangsi17}. Hence, the densest at-most-$k$-subgraph problem is also likely difficult to approximate. For the densest $k$  subgraph problem, the current best polynomial time approximation is $O(n^{1/4})$ \cite{BhaskaraCCFV10}.

\paragraph{Our results.} We show that, given a large partial solution to the densest subgraph or the densest at-most-$k$ subgraph problem, there exists a simple linear-time algorithm that produces a good approximation. 

For ease of notation, we use (i) $e(X)$ to denote the number of edges in the subgraph $G[X]$ induced by $X$, (ii) $e(X,Y)$ to denote the number of edges with one endpoint in $X$ and the other in $Y$, and (iii) $e(v,X)$ to denote the number of edges $vu$ such that $u \in X$.

Unless stated otherwise, $n$ and $m$ refer to the number of nodes and edges in the input graph respectively.

We first address the issue of uniqueness. Throughout, we let $H^\star$ denote a {\bf fixed} optimal solution, and the predictor predicts membership for $H^\star$. This assumption is often unavoidable in this setting, as is standard in the literature \cite{BravermanDSW24, AamandCGSW25}.

In real-world instances, the optimal solution is typically unique and, moreover, there are usually no \emph{competing} optimal solutions. That is, any solution whose objective value is close to optimal must substantially overlap with the optimal one. This type of \emph{stability} phenomenon has been studied extensively in problems such as clustering, independent set, and Max-Cut (see, e.g., \cite{BalcanHW20, bilu2012stable, AngelidakisABCD19, AwasthiBS12}).

However, for the densest subgraph problem, it is possible to enforce uniqueness as follows. If $H_1^\star$ and $H_2^\star$ are two solutions to the densest subgraph problem, then $H_1^\star \cup H_2^\star$ is also a solution to the densest subgraph problem. To see this, let $\Opt = \ds(H_1^\star) = \ds(H_2^\star)$. Note that $e(H_1^\star) = \Opt |H_1^\star|$, $e(H_2^\star) = \Opt |H_2^\star|$, and $e(H_1^\star \cap H_2^\star) \le \Opt |H_1^\star \cap H_2^\star|$.
\begin{align*}
    \ds(H_1^\star \cup H_2^\star) & = \frac{e(H_1^\star) + e(H_2^\star) - e(H_1^\star \cap H_2^\star)}{|H_1^\star| + |H_2^\star| - |H_1^\star \cap H_2^\star|} \\
    & \ge \frac{\Opt|H_1^\star| + \Opt|H_2^\star| - \Opt|H_1^\star \cap H_2^\star|}{|H_1^\star| + |H_2^\star| - |H_1^\star \cap H_2^\star|} = \Opt.
\end{align*}

Hence, we may enforce uniqueness for the densest subgraph problem by letting $H^\star$ be the maximal densest subgraph.

\begin{definition}
    We say that $S$ is a $(1-\epsilon)$ partial solution to $H^\star$ if the following holds:
    \begin{enumerate}
        \item $|S \cap H^{\star}| \geq (1- \epsilon)|H^{\star}|$ (i.e., few false negatives),
        \item $|S \setminus H^{\star}| \leq \epsilon|H^{\star}|$ (i.e., few false positives),
    \end{enumerate}
\end{definition}

This implies that the symmetric difference between $S$ and $H^\star$ is small $|H^\star \triangle S| \le 2 \epsilon |H^\star|$. Intuitively, the two conditions say that the provided partial solution $S$ has most nodes in $H^{\star}$ and few nodes outside $H^{\star}$. In the above definition, one can think of the predictor as a machine learning classifier that predicts whether $v \in H^\star$ based on its attributes. 

Next, we consider a scenario in which we have a predictor that predicts the probability that a node belongs to the densest subgraph $H^\star$. We use $o(v) \in [0,1]$ to denote the prediction for the probability that $v \in H^\star$ given the predictor. Some examples include logistic regression, neural networks with sigmoid or softmax outputs, naive Bayes classifier, and tree-based models with probability estimates.

\begin{definition}
    Let $g(v) = [v \in H^\star]$ denote the indicator variable for whether $v \in H^\star$. We say that $o$ is an $(\epsilon, \ell_p)$-probabilistic predictor for $H^\star$ if the following holds: for every node $v$, we are given a value $o(v) \in [0,1]$ such that
    \[
        \sum_{v \in V} \lvert o(v) - g(v) \rvert^p \le \epsilon \lvert H^\star \rvert.
    \]
\end{definition}

This condition covers important cases such as absolute error and squared error. 

\begin{theorem}[Main result 1]\label{thm:ds}
    Let $H^{\star} \subseteq V$ be the densest subgraph.
    \begin{enumerate}
        \item Given a $(1-\epsilon)$-partial solution $S$ to $H^{\star}$, there exists an $O(n+m)$-time algorithm that finds a subgraph whose density is at least 
        \[
            \left(1-O \left(\epsilon  + \frac{1}{|H^\star|} \right)\right) \ds(H^\star).
        \]
        \item Given an $(\epsilon, \ell_p)$-probabilistic predictor for $H^{\star}$, for some $p > 0$, there exists an $O(n+m)$-time algorithm that finds a subgraph whose density is at least 
        \[
            \left(1- 2^p \cdot O \left(\epsilon  + \frac{1}{|H^\star|} \right)\right) \ds(H^\star).
        \]
 
    \end{enumerate}
\end{theorem}

Furthermore, in linear time, for small enough $\epsilon$, this result is better than the 1/2-approximation of Charikar's  greedy algorithm \cite{Charikar00}. 

We first address the \textbf{robustness} of the above result. Since one can run both this algorithm and Charikar's greedy algorithm in linear $O(m+n)$ time and return the better solution, the worst-case approximation guarantee is therefore $1/2$. In terms of {\bf consistency}, as the predictor gets better (e.g., $\epsilon$ gets smaller), we get an approximation closer to $1$ without sacrificing the running time. 

One may argue that efficient algorithms and heuristics already exist for the densest subgraph problem and that the algorithm in Theorem \ref{thm:ds} is only interesting from a theoretical perspective. However, we want to point out that it can be extended to the problem of finding the densest at-most-$k$ subgraph. Even for $\epsilon$ that is not too small (say $\epsilon \approx 0.5$), we obtain a non-trivial constant approximation which is generally not possible assuming ETH.

This leads us to the second main result that addresses the NP-hard variant where we want to find the densest subgraph $H^\star$ subject to the constraint $|H^\star| \le k$.

\begin{theorem}[Main result 2]\label{thm:ds-amk}
    Let $H^{\star} \subseteq V$ be the densest subgraph with at most $k$ nodes. 
    \begin{enumerate}
        \item  Given a $(1-\epsilon)$-partial solution $S$ to $H^\star$,  there exists an $O(n+m)$-time algorithm that finds an at-most-$k$ subgraph whose density is at least 
        \[
            \left(1-O \left(\epsilon  + \frac{1}{|H^\star|} + \frac{1}{k}\right)\right) \ds(H^\star).
        \]
        \item Given an $(\epsilon, \ell_p)$-probabilistic predictor for $H^{\star}$, for some $p > 0$, there exists an $O(n+m)$-time algorithm that finds an at-most-$k$ subgraph whose density is at least 
        \[
            \left(1- 2^p \cdot O \left(\epsilon  + \frac{1}{|H^\star|} + \frac{1}{k}\right)\right) \ds(H^\star).
        \]
    \end{enumerate}
    
\end{theorem}

As this problem is NP-hard and it is likely that no constant-factor approximation exists in polynomial time, robustness is not guaranteed (even in the normal setting). As for consistency, as the prediction error $\epsilon$ decreases, we obtain an approximation ratio closer to $1$.

\paragraph{Application motivation.} In many applications, graphs are generated by the same underlying process, so patterns from past instances can help solve new ones more quickly.

Many real-world graphs also evolve continuously as new data arrives. Social networks gain or lose edges, and biological networks reconfigure over time. Detecting dense components in real time therefore requires repeatedly computing dense subgraphs on updated snapshots, which is important for tasks such as identifying emerging clusters. However, exact algorithms, especially for the densest at-most-$k$ subgraph problem, quickly become expensive.

In other settings, we analyze different parts of a large network that share common structure, such as research areas in citation networks, geographic regions in social networks, or pathways in biological networks. These subproblems are not independent: they often exhibit recurring structural patterns. A predictor trained on one region can therefore guide solutions in related regions. Our learning-augmented approach exploits this shared structure to achieve speedups while preserving near-optimal solutions.

\paragraph{Paper organization.} Algorithms and their analysis for the densest subgraph and densest at-most-$k$ subgraph are presented in Section \ref{sec:algorithms}. Section~\ref{sec:experiments} presents experiments of our algorithm for the densest at-most-$k$ subgraph problem on the Twitch Ego Nets and High-energy Physics Citation Network (HEP-PH) datasets.

\section{Algorithms}\label{sec:algorithms}
\subsection{Learning-Augmented Algorithms for Densest Subgraph} \label{sec:ds}
\paragraph{Hard Predictions.} This section demonstrates that, given a $1-\epsilon$ partial solution $S$ to the densest subgraph $H^\star$, we can compute a $1-\epsilon$ approximation in linear time. We propose a simple algorithm that augments $S$ by selecting an additional $\left\lceil \frac{\epsilon}{1 - \epsilon}|S| \right\rceil$ nodes from outside $S$ that have the highest number of connections to nodes in $S$. See Algorithm~\ref{alg:1} for the formal algorithm. A high-level justification of the algorithm’s correctness is illustrated in Figure~\ref{fig:undirected}.

\begin{figure}[ht]
    \centering
    \begin{minipage}{0.45\textwidth}
        \centering
        \begin{tikzpicture}[every node/.style={circle, draw, fill=black, inner sep=1.5pt}, node distance=1cm and 2cm]

        \node[draw=none, fill=none] at (-2.5,1.8) {\( H^\star \setminus S \)};
        \node[draw=none, fill=none] at (2.5,1.8) {\( H^\star \cap S \)};
        
        \coordinate (L1) at (-2,1);
        \coordinate (L2) at (-2,0.3);
        \coordinate (L3) at (-2,-0.4);
        \coordinate (L4) at (-2,-1.1);
        
        \coordinate (R1) at (2,1);
        \coordinate (R2) at (2,0.3);
        \coordinate (R3) at (2,-0.4);
        \coordinate (R4) at (2,-1.1);
        
        \draw (-2,0) ellipse (0.7cm and 1.5cm);
        \draw (2,0) ellipse (0.7cm and 1.5cm);
        
        \foreach \i in {1,2,3,4} {
            \node at (L\i) {};
            \node at (R\i) {};
        }
        
        \draw (L1) -- (R1);
        \draw (L1) -- (R2);
        \draw (L1) -- (R3);
        \draw (L2) -- (R2);
        \draw (L3) -- (R3);
        \draw (L4) -- (R3);
        \draw (L4) -- (R4);
        
        \end{tikzpicture}
    \end{minipage}%
    \hspace{1cm} 
    \begin{minipage}{0.45\textwidth}
        \caption{The approximation guarantee comes from the observation that the number of edges induced by $H^{\star} \setminus S$ is small, and the greedy approach ensures that we added at least as many edges to the solution as $|E(H^{\star} \setminus S, H^{\star} \cap S)|$.}
        \label{fig:undirected}
    \end{minipage}
\end{figure}

We first observe that the number of edges induced by the nodes of $H^\star$ outside $S$ is small.
\begin{proposition}\label{prop:false-negative-edges}
    Let $H^\star$ be the densest subgraph or the densest at-most-$k$ subgraph and $S$ be a $(1-\epsilon)$ partial solution to $H^\star$. Let $B = H^\star \setminus S$. Then, $e(B) \le \epsilon e(H^\star)$.
    \begin{proof}
        Recall that $|B| \le \epsilon |H^\star|$ according to the assumption on $S$. In the case of the densest subgraph problem, we have $\ds(B) \le \ds(H^\star)$. In the case of the densest at-most-$k$-subgraph problem, we also have $\ds(B) \le \ds(H^\star)$ because $|B| \le \epsilon |H^\star| \le k$ is a feasible solution.
        We have
        \begin{align*}
            \ds(B) & \le \ds(H^\star) \\
            \frac{e(B)}{|B|} & \le \frac{e(H^\star)}{|H^\star|} \\
            e(B) & \le \frac{e(H^\star)}{|H^\star|} |B| \le \epsilon e(H^\star). \qedhere
        \end{align*}     
    \end{proof}
\end{proposition}

\begin{algorithm}
    \caption{Approximate Densest Subgraph with Predictions}
    \label{alg:1}
    \KwIn{Graph $G = (V, E)$, and a set $S \subseteq V$ that is a $(1-\epsilon)$ partial solution to $H^\star$.}
    
    Let $U$ be the set of nodes outside $S$ of size $\min \left\{ \left\lceil \frac{\epsilon}{1-\epsilon}|S| \right\rceil , |V \setminus S| \right\}$  with highest $e(v,S)$ values. Ties broken arbitrarily.
    
    \Return $S \cup U$\;
\end{algorithm}

We now prove the first main result.

\begin{proof}[Proof of Theorem \ref{thm:ds} (Part 1)]
    Consider Algorithm~\ref{alg:1}. Let
    \[
    r := \min \left\{ \left\lceil \frac{\epsilon}{1-\epsilon}|S| \right\rceil,\ |V\setminus S| \right\}.
    \]

    Then $|U| = r$. We claim that $|U| \ge |H^\star \setminus S|$. Indeed,
    \begin{align*}
        \left\lceil \frac{\epsilon}{1-\epsilon}|S| \right\rceil  \ge \frac{\epsilon}{1-\epsilon}|S| & \ge \frac{\epsilon}{1-\epsilon}(1-\epsilon)|H^\star| \\
        &= \epsilon |H^\star| \ge |H^\star \setminus S|.
    \end{align*}

    The  inequality $|S|\ge (1-\epsilon)|H^\star|$ follows from the first property of $S$. Moreover, since $H^\star \setminus S \subseteq V\setminus S$, we also have
    $|V\setminus S| \ge |H^\star \setminus S|$. Thus, $r \ge |H^\star \setminus S|$.
   
     Note that we choose $U \subseteq V \setminus S$ in a way that maximizes $e(U,S)$ among all sets whose size is  $r \ge |H^\star \setminus S|$. Therefore,
    \begin{align}
        e(H^\star \cap S) + e(H^\star \setminus S, H^\star \cap S) + e(H^\star \setminus S) & = e(H^\star) \nonumber \\
        e(H^\star \cap S) + e(H^\star \setminus S, H^\star \cap S) & = e(H^\star) - e(H^\star \setminus S) \nonumber \\
        e(H^\star \cap S) + e(H^\star \setminus S, H^\star \cap S) & \ge (1-\epsilon) e(H^\star) \nonumber \\
        \implies e(H^\star \cap S) + e(H^\star \setminus S, S) & \ge (1-\epsilon) e(H^\star) \nonumber \\
        \implies e(S) + e(U,S) & \ge (1-\epsilon) e(H^\star) \nonumber \\
        \implies e(S \cup U) & \ge (1-\epsilon) e(H^\star). \label{eq:1}
    \end{align}
    The first inequality follows from Proposition~\ref{prop:false-negative-edges}. Hence,
    \begin{align*}
        \ds(S \cup U) & \ge (1-\epsilon) \frac{e(H^\star)}{|S \cup U|}.
    \end{align*}
    We also have
    \begin{align*}
        |S \cup U| 
        & \le |S| + \left\lceil \frac{\epsilon}{1-\epsilon}|S| \right\rceil \\
        & \le |S| + \frac{\epsilon}{1-\epsilon}|S| + 1 \\
        & \le \left(1 + \epsilon + \frac{\epsilon}{1-\epsilon}\right)|H^\star| + 1.
    \end{align*}
    Therefore,
    \begin{align*}
        \ds(S \cup U)
        & = \frac{e(S \cup U)}{|S \cup U|} \\
        & \ge \frac{(1-\epsilon)e(H^\star)}{(1+\epsilon + \epsilon/(1-\epsilon))|H^\star| + 1} \\
        & = \frac{1-\epsilon}{1+\epsilon + \epsilon/(1-\epsilon) + 1/|H^\star|} \ds(H^\star) \\
        & =\left(1-O \left(\epsilon  + \frac{1}{|H^\star|} \right)\right) \ds(H^\star).
    \end{align*}
    
    It is straightforward to implement this algorithm in $O(m+n)$ time by iterating over each edge $uv$ and, if $u \notin S$ and $v \in S$ (using a hash table to check membership in $S$), incrementing $e(u,S)$. We then use linear-time selection to find $U$.
\end{proof}

\paragraph{Probabilistic Predictions.}

Next, we consider a scenario in which we have an $(\epsilon, \ell_p)$ probabilistic predictor for $H^\star$. We also want to design a fast algorithm that achieves a good approximation in this case. In particular, we will show that it is possible to obtain a $1-O(2^p \cdot \epsilon)$ approximation. 

Let $\delta$ be a value to be determined later. We define 
\[
    S := \{ v \in V: o(v) \ge 1 - \delta \}.
\]

We rely on the following technical lemmas. 

\begin{lemma}\label{lem:prob-1}
    Let $o$ be an $(\epsilon, \ell_p)$ probabilistic predictor for $H$, where $H$ is an arbitrary subgraph.  Let $S = \{ v \in V: o(v) \ge 1 - \delta \}$. We have
    \[
        |H \cap S| \ge |H| \left( 1-\frac{ \epsilon}{\delta^p} \right).
    \]
\end{lemma}
\begin{proof}
    Let $g(v) = [v\in H]$ be the indicator variable for the event $v \in H$. Note that
    \begin{align*}
         \sum_{v \in H}  (1 - o(v))^p = \sum_{v \in H}  |o(v) - g(v)|^p \le  \sum_{v \in V} |o(v) - g(v)|^p \le \epsilon |H| .
    \end{align*}
    
    The last inequality above follows from the definition of $o$.  For any $v \in H \setminus S$, we have
    \begin{align*}
        o(v) & < 1 - \delta  \\
         \delta^p & < (1 - o(v))^p \\
        \sum_{v \in H \setminus S} \delta^p & < \sum_{v \in H \setminus S} (1-o(v))^p \\
        |H \setminus S| & < \frac{\epsilon |H|}{\delta^p}. \qedhere
    \end{align*}
\end{proof}

\begin{lemma}\label{lem:prob-2}
    Let $o$ be an $(\epsilon, \ell_p)$ probabilistic predictor for $H$, where $H$ is an arbitrary subgraph. Let $S = \{ v \in V: o(v) \ge 1 - \delta \}$. We have 
    \[
        |S \setminus H| \le \frac{|H| \epsilon}{(1-\delta)^p}.
    \]
\end{lemma}

\begin{proof}
    Again, let $g(v) = [v\in H]$. For any $v \in S \setminus H$,
    \begin{align*}
        1 - \delta & \le o(v) \\
        (1-\delta)^p & \le o(v)^p \\
        \sum_{v \in S \setminus H} (1-\delta)^p & \le \sum_{v \in S \setminus H} o(v)^p \\
        |S \setminus H| & \le  \frac{\sum_{v \in S \setminus H} o(v)^p}{(1-\delta)^p}~.
    \end{align*}
    Observe that 
    \[
        \sum_{v \in S \setminus H} o(v)^p \le \sum_{v \notin H} o(v)^p = \sum_{v \notin H} |o(v) - g(v)|^p \le \epsilon |H|.
    \]
    Hence, we deduce that  
    \[
        |S \setminus H | \le \frac{\epsilon |H|}{(1-\delta)^p}.\qedhere
    \]
\end{proof}

We combine the above as a lemma saying that one can obtain a $(1- \epsilon \cdot 2^p)$ partial solution given an $(\epsilon, \ell_p)$ probabilistic predictor.

\begin{lemma}\label{lem:predictor-conversion}
    Given a $(\epsilon, \ell_p)$ probabilistic predictor for $H$, we can obtain a $(1-2^p \epsilon)$ partial solution to $H$ in $O(n)$ time.
\end{lemma}

\begin{proof}
    Applying Lemma \ref{lem:prob-1} and Lemma \ref{lem:prob-2} with $\delta = 1/2$, we have a set $S$ such that
    \begin{align*}
        |S \setminus H| \le 2^p \epsilon |H|  \\
        |H \cap S| \ge  \left( 1-2^p \epsilon \right) |H|.
    \end{align*}
    Therefore, $S$ is a $(1-2^p \epsilon)$ partial solution to $H$.
\end{proof}

Putting the above together, we get the second main result.

\begin{proof}[Proof of Theorem \ref{thm:ds} (Part 2)]
    This follows directly from Lemma \ref{lem:predictor-conversion} and Theorem \ref{thm:ds} (Part 1).
\end{proof}

\subsection[Learning-Augmented Algorithm for Densest At-Most-k-Subgraph]{Learning-Augmented Algorithm for Densest At-Most-$k$-Subgraph} \label{sec:ds-amk}

In this section, we present a fast algorithm that approximates the densest subgraph with at most $k$ nodes.
To begin, we restate our assumption about the predictor. Let $H^\star$ be the densest at-most-$k$-subgraph and $S$ be a $(1-\epsilon)$ partial solution to $H^\star$.

We consider the following algorithm, which first adds $\left\lceil \frac{\epsilon}{1-\epsilon}|S| \right\rceil$ nodes $U$ that have the highest number of connections to the predicted solution $S$. Then, starting with $T = U \cup S$, it iteratively removes nodes from $T$ that currently have the lowest degree in $T$ so that the final solution has at most $k$ (trimming step). 

\begin{algorithm}
\caption{Densest at-most-$k$ Subgraph from Predictions}
\label{alg:amk}
\KwIn{Graph $G=(V,E)$, $S \subseteq V$ that is a $(1-\epsilon)$ partial solution to $H^\star$ and $k \in \mathbb{Z}^+$}
\KwOut{Set $T \subseteq V$ with $|T| \le k$}

Let $r \gets \min \left \{ \left\lceil \frac{\epsilon}{1-\epsilon}|S| \right\rceil, |V \setminus S| \right \}$;

Let $U \subseteq V \setminus S$ be the $r$ nodes with the largest values of $e(v,S)$\;

$T \gets S \cup U$\;

\tcp{Trimming step}
\While{$|T| > k$}{
    remove from $T$ a node of minimum degree in $G[T]$\;
}

\Return{$T$}
\end{algorithm}

Let $T_0 = U \cup S$. The only extra argument that we need to make is that iteratively removing nodes with the lowest degree in $T$ (to satisfy the ``at most $k$'' constraint) does not lower the density too much. 

\begin{lemma}\label{lem:iterative-removal}
Let $T_0$ be a graph on $y$ nodes. For $i \ge 1$, let $T_i$ be the graph after removing the node with the lowest degree from $T_{i-1}$. Then,
\[
\ds(T_r) \;\ge\; \frac{y-r-1}{y-1}\,\ds(T_0).
\]
\end{lemma}

\begin{proof}
Let $m_i$ and $n_i$ be the number of edges and nodes in $T_i$, respectively. First, note that $\ds(T_0) = \frac{m_0}{y}$.

Consider any intermediate graph $T_i$ with $n_i := y-i$ vertices and $m_i$ edges. Let $\delta$ denote the minimum degree in $T_i$. Since the average degree in $T_i$ is $2m_i/n_i$, we have
\[
\delta \le \frac{2m_i}{n_i}.
\]
Removing a node of degree $\delta$ deletes exactly $\delta$ edges, so
\[
m_{i+1} = m_i - \delta \ge m_i - \frac{2m_i}{n_i}.
\]
The density of $T_{i+1}$ therefore satisfies
\begin{align*}
\ds(T_{i+1})
&= \frac{m_{i+1}}{n_i-1}
\ge \frac{m_i - \frac{2m_i}{n_i}}{n_i-1}
= \frac{m_i}{n_i} \cdot \frac{n_i-2}{n_i-1}
= \ds(T_i) \cdot \frac{n_i-2}{n_i-1}.
\end{align*}

Applying this inequality iteratively for $r$ steps starting from $T_0$, we obtain
\begin{align*}
    \ds(T_r) &\ge \ds(T_0)
\prod_{j=0}^{r-1} \frac{(y-j)-2}{(y-j)-1} \\
& = \ds(T_0) \frac{n_0 - 2}{n_0 - 1} \cdot \frac{n_0 - 3}{n_0 - 2} \cdot \frac{n_0 - 4}{n_0 - 3} \cdot \ldots \cdot \frac{n_0 - (r+1)}{n_0 - r}.
\end{align*}

This product telescopes, yielding
\[
\ds(T_r) \ge \ds(T_0)\cdot \frac{y-r-1}{y-1}. \qedhere
\]
\end{proof}

We now put together the details to prove Theorem~\ref{thm:ds-amk}.

\begin{proof}[Proof of Theorem \ref{thm:ds-amk}]

We use the same argument as in the proof of Theorem~\ref{thm:ds} (Part~1) to get
\[
|T_0| = |S \cup U|
\le |S| + \left\lceil \frac{\epsilon}{1-\epsilon}|S| \right\rceil
\le (1+\epsilon + \epsilon/(1-\epsilon))|H^\star| + 1.
\]
and
\begin{align*}
    \ds(T_0)
    = \frac{e(S \cup U)}{|S \cup U|}
    \ge \frac{(1-\epsilon)e(H^\star)}{(1+\epsilon + \epsilon/(1-\epsilon))|H^\star| + 1}.
\end{align*}

In Algorithm~\ref{alg:amk}, the final output $T$ has at most $k$ nodes. If no node is removed from $T_0$, then $T=T_0$ and we are done. Otherwise, applying Lemma ~\ref{lem:iterative-removal} with $r = |T_0| - k$ yields
\begin{align*}
    \ds(T)
    &\ge \frac{|T_0| - (|T_0|-k) - 1}{|T_0|-1}\ds(T_0) \\
    &\ge \frac{k-1}{|T_0|-1}\frac{(1-\epsilon)e(H^\star)}{(1+\epsilon + \epsilon/(1-\epsilon))|H^\star| + 1}.
\end{align*}
As $|T_0| \le (1+\epsilon + \epsilon/(1-\epsilon))k + 1$, this implies
\begin{align*}
    \ds(T) \ge  & \frac{k-1}{ (1+\epsilon + \epsilon/(1-\epsilon))k} \cdot \frac{(1-\epsilon)e(H^\star)}{(1+\epsilon + \epsilon/(1-\epsilon))|H^\star| + 1}\\
    & =  \frac{(1-1/k)}{ 1+\epsilon + \epsilon/(1-\epsilon)}  \cdot \frac{(1-\epsilon)}{1+\epsilon + \epsilon/(1-\epsilon) + 1/|H^\star|} \ds(H^\star) \\
    & = \left(1-O \left(\epsilon + \frac{1}{k} + \frac{1}{|H^\star|} \right)\right) \cdot \ds(H^\star)
\end{align*}
    
In terms of running time, the only difference compared to the densest
subgraph algorithm is the additional trimming step, which can be
implemented in $O(m+n)$ time using standard techniques as in
\cite{Charikar00}. 

Since the degree of any vertex is an integer in the range
$\{0,1,\dots,n\}$, we maintain an array of buckets
$B[0], B[1], \dots, B[n]$, where each bucket is implemented as a
doubly-linked list.
Bucket $B[i]$ stores exactly the vertices whose current degree is $i$
in the subgraph induced by the remaining vertices.
For each vertex $v$, we additionally maintain a pointer to its position
in the corresponding bucket, allowing deletions and insertions to be
performed in constant time.

In each iteration, the algorithm removes a vertex from the nonempty
bucket with minimum index, corresponding to a minimum-degree vertex.
When a vertex $v$ is removed, we examine all neighbors $u$ of $v$ that
remain in the graph.
For each such neighbor, the degree of $u$ decreases by one, and $u$ is
moved from bucket $B[\deg(u)]$ to bucket $B[\deg(u)-1]$.
Because the buckets are doubly linked lists and each vertex stores a
pointer to its list node, each such move can be carried out in $O(1)$
time.
Each edge is responsible for at most one degree decrement, and therefore
the total time spent updating degrees over the entire execution is
$O(n+m)$.

Let $d$ denote the current minimum degree.
Since $d$ can be increased at most $n$ times and each vertex is deleted
exactly once, the overall running time of the algorithm is $O(m+n)$.

The second part follows directly from Lemma~\ref{lem:predictor-conversion}.
\end{proof}

\section{Experiments}\label{sec:experiments}
\subsection{Experimental Setup}

\paragraph{Setup.} We run all experiments on a machine with Linux Mint 22.3, an AMD Ryzen 9900X CPU with 24 threads, and 64~GB of DDR5 RAM. All experiments are implemented in Julia.

Because existing algorithms for the densest subgraph problem are already quite efficient in practice, we focus on the densest at-most-$k$-subgraph problem. We use two datasets: Twitch Ego Nets and the High-Energy Physics Citation Network from \cite{snapnets}.

For this problem, we focus on regimes in which $k$ is not too small relative to $n$. When $k$ is very small compared to $n$, the labels become too imbalanced for a predictor to perform well.

\paragraph{Pruning.} In order to train a predictor for the densest at-most-$k$-subgraph problem, we need to compute the exact solution for some training graphs to obtain ground truth labels. We propose some simple pruning rules that reduce the search space.

Specifically, we iteratively remove nodes with minimum degree from the graph until only $k$ nodes remain. Let the resulting graph be denoted by $H$, and let $d' = \ds(H)$. Starting from $H$, we again iteratively remove nodes of minimum degree and update the maximum density attained by any intermediate subgraph during this process in $d'$ and $H$. 

Since $H$ is a subgraph with at most $k$ nodes, $d'$ serves as a lower bound on the density of the densest subgraph with at most $k$ nodes.

We then iteratively remove nodes in the original graph $G$ whose current degree is strictly smaller than $d'$ until it is no longer possible to do so. We claim that any node removed by this procedure cannot belong to an optimal solution. To see this, suppose that $v \in H^\star$ is the first node in $H^\star$ to be removed, and let $R$ denote the remaining graph at the moment $v$ is removed. We have
\begin{align*}
    \ds(H^\star \setminus \{v\}) 
    &= \frac{e(H^\star) - e(v, H^\star)}{|H^\star| - 1} \\
    &\ge \frac{e(H^\star) - e(v, R)}{|H^\star| - 1} \\
    &> \frac{e(H^\star) - \ds(H^\star)}{|H^\star| - 1} \\
    &= \frac{e(H^\star) - e(H^\star)/|H^\star|}{|H^\star| - 1} = \frac{e(H^\star)}{|H^\star|}.
\end{align*}
This is a contradiction, since $H^\star \setminus \{v\}$ is also a feasible solution and achieves strictly higher density than $H^\star$. Therefore, the removed nodes cannot be part of an optimal solution. We can then repeat the procedure on the remaining graph until no more nodes can be removed. This pruning rule allows us to significantly speed up our exact algorithms. 

We also implemented several other pruning techniques. Let $d'$ be the lower bound for the optimal density obtained by the pruning procedure described above.  We know that any subset of vertices of size smaller than $2 d' + 1$ cannot have a density higher than $d'$; therefore, we can safely ignore all subsets of vertices of size smaller than $2 d' + 1$.

\subsection{Twitch Ego Nets}

The first dataset we use consists of ego-networks of Twitch users who participated in the partnership program in April 2018, as described in \cite{karateclub}. It contains over 127,000 graphs. Each node represents a user, and each edge represents a friendship. 

We restrict our experiment to graph with $n \le 100$ and use $k=15$. 

\begin{figure}
    \centering
    \includegraphics[width=0.4\linewidth]{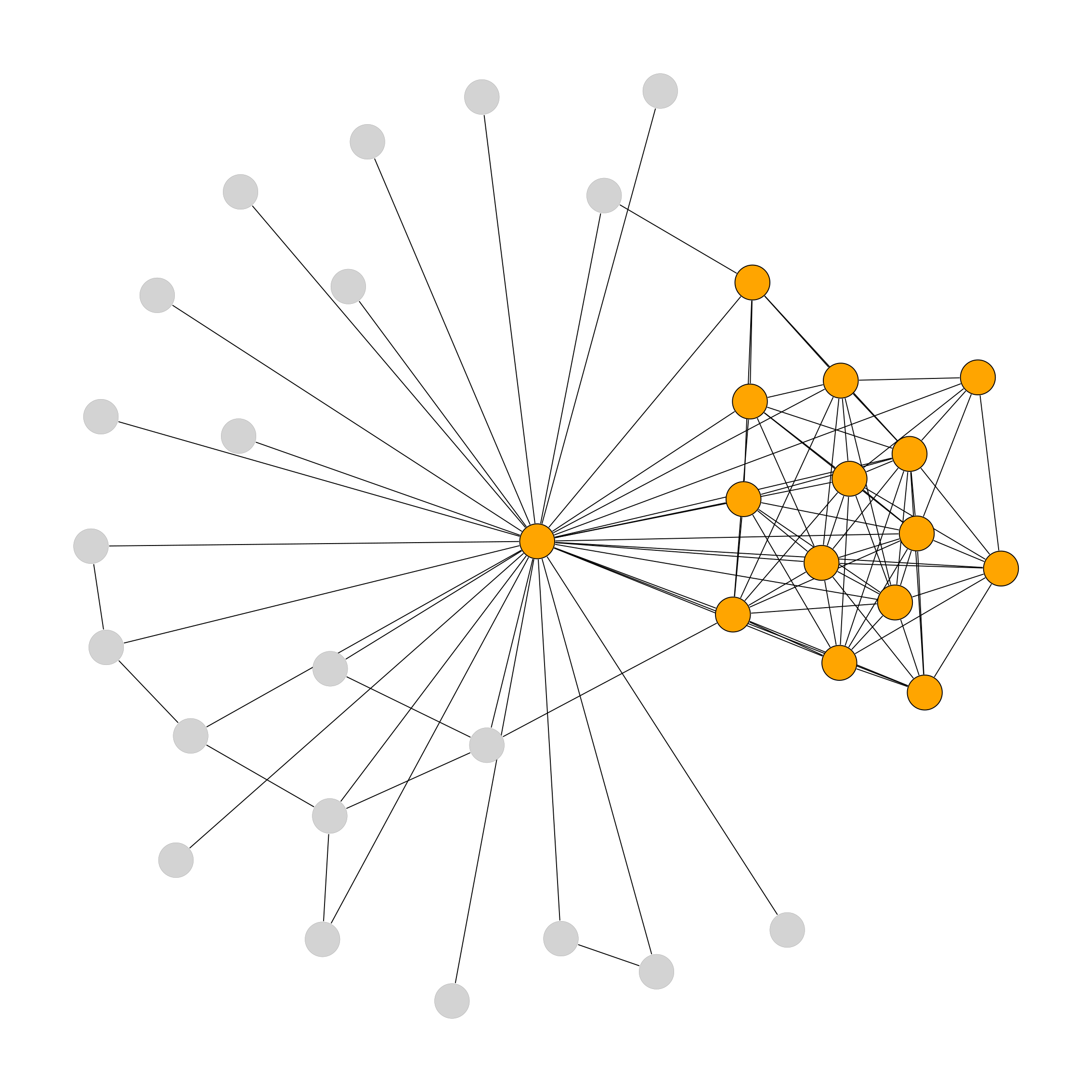}
    \caption{An example of a densest at-most-15 densest subgraph in the Twitch ego data set.}
    \label{fig:damks_example}
\end{figure}

Since the nodes lack attributes, we follow standard practice in graph learning and use local structural features. Specifically, each node is characterized by its degree, the average degree of its neighbors, and the total number of nodes in its graph. These features can be computed for all nodes in $O(n+m)$ time, matching the linear-time complexity of our algorithm. The label is $1$ if a node belongs to the densest at-most-$15$ subgraph, and $0$ otherwise. We obtain the ground truth by brute-force, using the pruning techniques described below.

We use a set of 200 graphs for training and testing, and another 200 graphs for validation. These graphs are chosen uniformly at random. Some instances are dropped because computing the ground truth times out. In total, we retain 179 graphs for training and testing and 173 graphs for validation.

We train our predictor using a random forest classifier with 10 decision trees. For the 179 graphs in the training/testing set, we perform an 80/20 split at the node level: 80\% of the nodes are used for training and the remaining 20\% for testing. On the testing set, the confusion matrix is shown Table \ref{tab:confusion-matrix}.

\begin{table}[h]
\centering
\begin{tabular}{|c|c|c|}
\hline
 & \textbf{Predicted 0} & \textbf{Predicted 1} \\
\hline
\textbf{Actual 0} & 527 & 83 \\
\hline
\textbf{Actual 1} & 91 & 319 \\
\hline
\end{tabular}
\caption{Confusion matrix for the random forest classifier predicting whether a node belongs to the densest at-most-15 subgraph in the Twitch ego data set.}
\label{tab:confusion-matrix}
\end{table}

This suggests a false positive rate of 13.6\% and a false negative rate of 22.2\%. Hence, we set $\epsilon = 0.2$.

\begin{figure}
    \centering

    \includegraphics[width=0.4\linewidth]{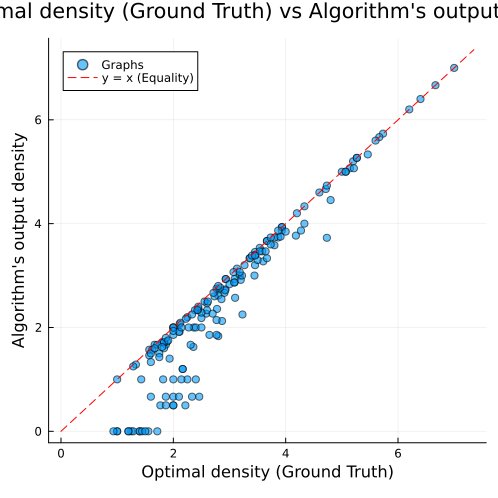}
    \includegraphics[width=0.4\linewidth]{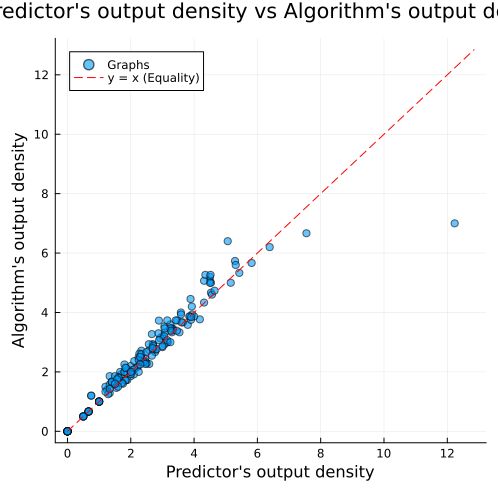}

    \caption{Pairwise comparison among our algorithm's solution (Section \ref{sec:ds-amk}), the optimal solution, and the density of the predicted set of nodes for the densest at-most-$15$ subgraph problem in the Twitch ego data set.}
    \label{fig:damks-scatter}
\end{figure}

We compare the density obtained by our algorithm in Section \ref{sec:ds-amk}, the optimal solution, and the density of the predictor's solution (i.e., the nodes that are predicted to be in the densest at-most-15 subgraph by the random forest). We emphasize that the predictor's solution might have more than $k$ nodes and may not be a feasible solution. See Figure \ref{fig:damks-scatter}. The average and median approximation factors of our algorithm in the validation set are 0.81 and 0.94 respectively.

\begin{table}[h]
\centering
\begin{tabular}{|l|c|}
\hline
\textbf{Metric} & \textbf{Value} \\
\hline
Average Approximation Ratio & 0.806 \\
Median Approximation Ratio & 0.943 \\
\hline
\textbf{High Quality Solutions ($\ge 0.95$)} & \textbf{81 / 173 (46.82\%)} \\
\hline
\end{tabular}
\caption{Approximation factors of Algorithm \ref{alg:amk} on validation graphs for the densest at-most-$15$ subgraph problem }
\label{tab:validation-approx-ratio}
\end{table}

We also compare the time to compute the exact solution using brute-force with pruning techniques and the learning-augmented algorithm described in Section \ref{sec:ds-amk} over the validation set. The total runtime of the learning-augmented algorithm over all 173 validation graphs is $2.1$ seconds, compared to $374.52$ seconds for the exact algorithm (a speedup of over $178\times$ on average).

\begin{table}[h]
\centering
\begin{tabular}{|l|c|}
\hline
\textbf{Algorithm} & \textbf{Runtime} \\
\hline
Learning Augmented algorithm  & 2.1 s \\
\hline
Exact algorithm (brute force with pruning) & 374.52 s \\
\hline
\end{tabular}
\caption{Runtime comparison between the learning-augmented algorithm (Section \ref{sec:ds-amk}) and the exact algorithm on validation graphs for the densest at-most-$15$ subgraph problem over 173 graphs in the Twitch ego data set's validation set.}
\label{tab:runtime-comparison}
\end{table}

\begin{figure}[H]
    \centering
    \includegraphics[scale=0.3]{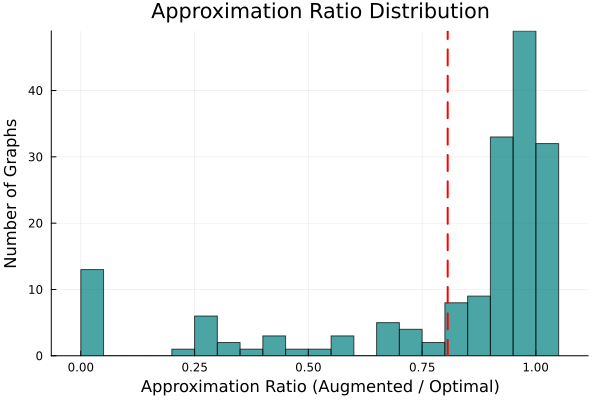}
    \caption{Approximation factors of Algorithm \ref{alg:amk} on  testing graphs for the densest at-most-$15$ subgraph problem in the Twitch ego data set.}
    \label{fig:damks-approx}
\end{figure}

\subsection{High-energy Physics Citation Network}

The arXiv HEP-PH citation graph covers all citations among 34,546 high energy physics phenomenology papers, comprising 421,578 edges. Here, we randomly sample induced subgraphs of size 50 nodes by picking a random node and start a breadth-first search from it until we hit the desired number of nodes.  For this experiment, we set $k=10$.

This dataset is significantly more challenging since the graphs are denser everywhere and it is harder to find the exact solution. We had to filter out some hard instances where the exact solution cannot be obtained within a reasonable time frame. In total, the number of graphs for training and testing is 91 and the number of graphs for validation is 96.

\begin{figure}[h]
    \centering
    \includegraphics[width=0.5\linewidth]{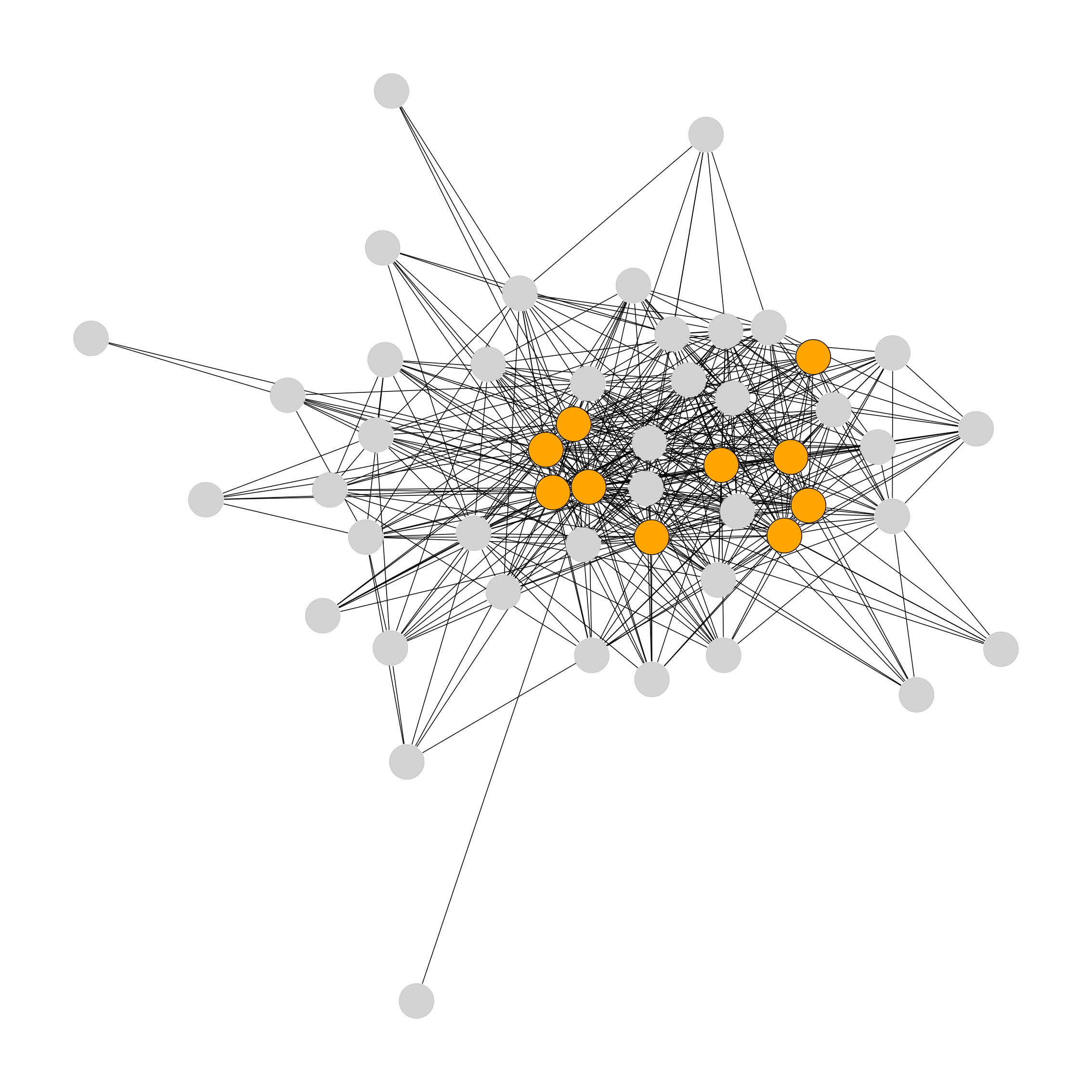}
    \caption{An example of the densest at-most-$10$ subgraph in the HEP-PH citation network.}
    \label{fig:damks-example-2}
\end{figure}

We again train a random forest with 10 decision trees as the classifier using the same features as in the Twitch ego data set. The confusion matrix is as follows:

\begin{table}[h]
\centering
\begin{tabular}{|c|c|c|}
\hline
 & \textbf{Predicted 0} & \textbf{Predicted 1} \\
\hline
\textbf{Actual 0} & 548 & 47 \\
\hline
\textbf{Actual 1} & 93 & 94 \\
\hline
\end{tabular}
\caption{Confusion matrix of the random forest predictor on the HEP-PH dataset.}
\label{tab:confusion-hepph}
\end{table}

From this confusion matrix, we choose $\epsilon = 0.497$. Even though the predictor is less accurate than the one for the Twitch ego data set, our learning-augmented algorithm still achieves a respectable average and median approximation factor of 0.81 and 0.93 respectively. Recall that this problem is presumably hard to approximate in general. 

\begin{figure}
    \centering

    \includegraphics[width=0.4\linewidth]{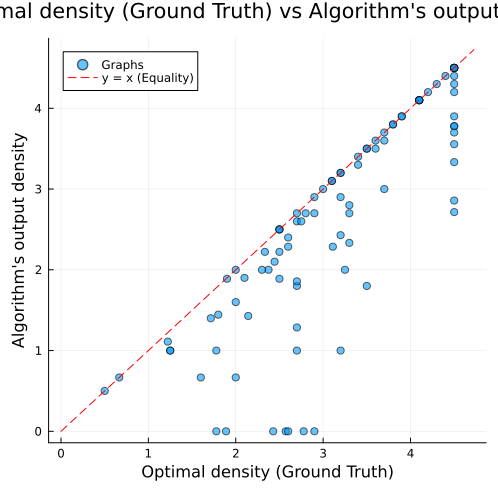}
    \includegraphics[width=0.4\linewidth]{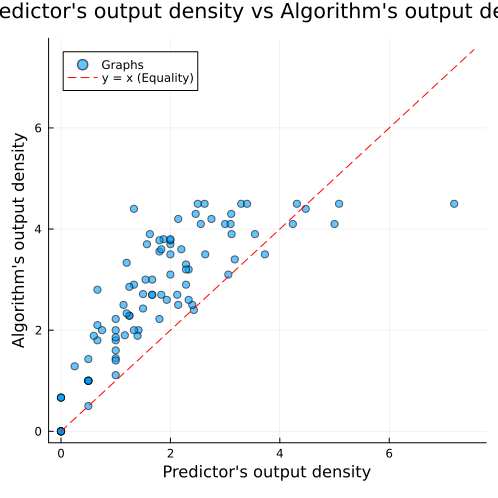}

    \caption{Pairwise density comparison on the HEP-PH dataset: augmented algorithm vs.\ optimal (left) and predictor vs.\ augmented algorithm (right), for the densest at-most-$k$ subgraph problem.}
    \label{fig:damks-scatter-hepph}
\end{figure}

Figure \ref{fig:damks-scatter-hepph} shows how important the learning-augmented algorithm is compared to the predictor's solution. Not only does it makes the solution feasible, but it also significantly improves the density compared to the predictor's solution. 

\begin{table}[h]
\centering
\begin{tabular}{|l|c|}
\hline
\textbf{Metric} & \textbf{Value} \\
\hline
Average Approximation Ratio & 0.8097 \\
Median Approximation Ratio & 0.9322 \\
\hline
\textbf{High Quality Solutions ($\ge 0.95$)} & \textbf{46 / 96 (47.92\%)} \\
\hline
\end{tabular}
\caption{Approximation factors of the learning-augmented algorithm on the HEP-PH validation graphs.}
\label{tab:approx-hepph}
\end{table}

\begin{figure}[H]
    \centering
    \includegraphics[scale=0.3]{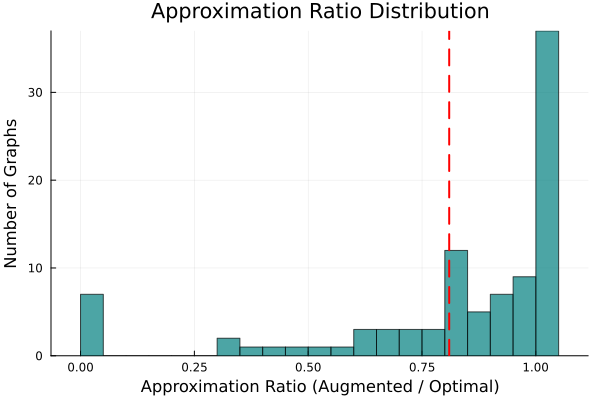}
    \caption{Distribution of approximation ratios (augmented / optimal) on the HEP-PH validation graphs.}
    \label{fig:damks-approx-hepph}
\end{figure}

The total runtime of the learning-augmented algorithm over all 96 validation graphs is shown in Table \ref{tab:runtime-hepph}. 

\begin{table}[h]
\centering
\begin{tabular}{|l|c|c|c|}
\hline
\textbf{Algorithm} & \textbf{Mean} & \textbf{Median} & \textbf{Total} \\
\hline
Augmented & 22.43 ms & 0.01 ms & 2.15 s \\
Exact (brute-force with pruning) & 14305.8 ms & 2223.39 ms & 1373.36 s \\
\hline
\end{tabular}
\caption{Runtime comparison on the HEP-PH validation graphs (96 instances) between the learning-augmented algorithm and the exact brute-force algorithm.}
\label{tab:runtime-hepph}
\end{table}

\section{Conclusion}

In this paper, we presented several simple and practical learning-augmented linear-time algorithms for computing the densest subgraph and its variants, including the NP-hard densest at-most-$k$ subgraph. These algorithms offer several desirable properties. They come with provable approximation guarantees and incur minimal computational overhead, running in linear time aside from the cost of executing the predictor. We also demonstrated, both theoretically and empirically, that blindly relying on the predictor can lead to poor approximations or infeasible solutions. When used appropriately, our learning-augmented algorithms provide a practical way to obtain a high-quality solution to the densest at-most-$k$ subgraph problem.

\bibliographystyle{plain}
\bibliography{references}

\begin{thebibliography}{10}

\bibitem{algorithms_with_predictions}
Algorithms with predictions.
\newblock \url{https://algorithms-with-predictions.github.io/}.
\newblock Accessed: 2025-04-15.

\bibitem{AamandCGSW25}
Anders Aamand, Justin~Y. Chen, Siddharth Gollapudi, Sandeep Silwal, and Hao Wu.
\newblock Improved approximations for hard graph problems using predictions.
\newblock In {\em {ICML}}, Proceedings of Machine Learning Research. {PMLR} / OpenReview.net, 2025.

\bibitem{Andersen}
Reid Andersen.
\newblock Finding large and small dense subgraphs.
\newblock {\em CoRR}, abs/cs/0702032, 2007.

\bibitem{AndersenC09}
Reid Andersen and Kumar Chellapilla.
\newblock Finding dense subgraphs with size bounds.
\newblock In {\em {WAW}}, volume 5427 of {\em Lecture Notes in Computer Science}, pages 25--37. Springer, 2009.

\bibitem{AngelidakisABCD19}
Haris Angelidakis, Pranjal Awasthi, Avrim Blum, Vaggos Chatziafratis, and Chen Dan.
\newblock Bilu-linial stability, certified algorithms and the independent set problem.
\newblock In {\em {ESA}}, volume 144 of {\em LIPIcs}, pages 7:1--7:16. Schloss Dagstuhl - Leibniz-Zentrum f{\"{u}}r Informatik, 2019.

\bibitem{AwasthiBS12}
Pranjal Awasthi, Avrim Blum, and Or~Sheffet.
\newblock Center-based clustering under perturbation stability.
\newblock {\em Inf. Process. Lett.}, 112(1-2):49--54, 2012.

\bibitem{BahmaniGM14}
Bahman Bahmani, Ashish Goel, and Kamesh Munagala.
\newblock Efficient primal-dual graph algorithms for mapreduce.
\newblock In {\em {WAW}}, volume 8882 of {\em Lecture Notes in Computer Science}, pages 59--78. Springer, 2014.

\bibitem{BahmaniKV12}
Bahman Bahmani, Ravi Kumar, and Sergei Vassilvitskii.
\newblock Densest subgraph in streaming and mapreduce.
\newblock {\em {PVLDB}}, 5(5):454--465, 2012.

\bibitem{BalcanHW20}
Maria{-}Florina Balcan, Nika Haghtalab, and Colin White.
\newblock \emph{k}-center clustering under perturbation resilience.
\newblock {\em {ACM} Trans. Algorithms}, 16(2):22:1--22:39, 2020.

\bibitem{BXGPF13}
Alex Beutel, Wanhong Xu, Venkatesan Guruswami, Christopher Palow, and Christos Faloutsos.
\newblock Copycatch: Stopping group attacks by spotting lockstep behavior in social networks.
\newblock In {\em Proceedings of the 22Nd International Conference on World Wide Web (WWW)}, pages 119--130, 2013.

\bibitem{BhaskaraCCFV10}
Aditya Bhaskara, Moses Charikar, Eden Chlamtac, Uriel Feige, and Aravindan Vijayaraghavan.
\newblock Detecting high log-densities: an \emph{O}(\emph{n}\({}^{\mbox{1/4}}\)) approximation for densest \emph{k}-subgraph.
\newblock In {\em {STOC}}, pages 201--210. {ACM}, 2010.

\bibitem{BHNT15}
Sayan Bhattacharya, Monika Henzinger, Danupon Nanongkai, and Charalampos~E. Tsourakakis.
\newblock Space- and time-efficient algorithm for maintaining dense subgraphs on one-pass dynamic streams.
\newblock In {\em Proc.~47th {ACM} Symposium on Theory of Computing ({STOC})}, pages 173--182, 2015.

\bibitem{bilu2012stable}
Yonatan Bilu and Nathan Linial.
\newblock Are stable instances easy?
\newblock {\em Combinatorics, Probability and Computing}, 21(5):643--660, 2012.

\bibitem{BoldrinV24}
Cristian Boldrin and Fabio Vandin.
\newblock Fast and accurate triangle counting in graph streams using predictions.
\newblock In {\em {ICDM}}, pages 31--40. {IEEE}, 2024.

\bibitem{BoobGPSTWW20}
Digvijay Boob, Yu~Gao, Richard Peng, Saurabh Sawlani, Charalampos~E. Tsourakakis, Di~Wang, and Junxing Wang.
\newblock Flowless: Extracting densest subgraphs without flow computations.
\newblock In {\em {WWW}}, pages 573--583. {ACM} / {IW3C2}, 2020.

\bibitem{BravermanDSW24}
Vladimir Braverman, Prathamesh Dharangutte, Vihan Shah, and Chen Wang.
\newblock Learning-augmented maximum independent set.
\newblock In {\em {APPROX/RANDOM}}, volume 317 of {\em LIPIcs}, pages 24:1--24:18. Schloss Dagstuhl - Leibniz-Zentrum f{\"{u}}r Informatik, 2024.

\bibitem{Charikar00}
Moses Charikar.
\newblock Greedy approximation algorithms for finding dense components in a graph.
\newblock In {\em {APPROX}}, volume 1913 of {\em Lecture Notes in Computer Science}, pages 84--95. Springer, 2000.

\bibitem{ChekuriQT22}
Chandra Chekuri, Kent Quanrud, and Manuel~R. Torres.
\newblock Densest subgraph: Supermodularity, iterative peeling, and flow.
\newblock In {\em {SODA}}, pages 1531--1555. {SIAM}, 2022.

\bibitem{CS12}
Jie Chen and Yousef Saad.
\newblock Dense subgraph extraction with application to community detection.
\newblock {\em IEEE Trans. on Knowl. and Data Eng.}, 24(7):1216--1230, 2012.

\bibitem{ChenSVZ22}
Justin~Y. Chen, Sandeep Silwal, Ali Vakilian, and Fred Zhang.
\newblock Faster fundamental graph algorithms via learned predictions.
\newblock In {\em {ICML}}, volume 162 of {\em Proceedings of Machine Learning Research}, pages 3583--3602. {PMLR}, 2022.

\bibitem{Dong0V25}
Yinhao Dong, Pan Peng, and Ali Vakilian.
\newblock Learning-augmented streaming algorithms for approximating {MAX-CUT}.
\newblock In {\em {ITCS}}, volume 325 of {\em LIPIcs}, pages 44:1--44:24. Schloss Dagstuhl - Leibniz-Zentrum f{\"{u}}r Informatik, 2025.

\bibitem{DGP07}
Yon Dourisboure, Filippo Geraci, and Marco Pellegrini.
\newblock Extraction and classification of dense communities in the web.
\newblock In {\em Proc.~16th International Conference on World Wide Web (WWW)}, pages 461--470, 2007.

\bibitem{EsfandiariHW15}
Hossein Esfandiari, MohammadTaghi Hajiaghayi, and David~P. Woodruff.
\newblock Applications of uniform sampling: Densest subgraph and beyond.
\newblock {\em CoRR}, abs/1506.04505, 2015.

\bibitem{FeijenGuido2021}
Willem Feijen and Guido Sch{\"{a}}fer.
\newblock Using machine learning predictions to speed-up dijkstra's shortest path algorithm.
\newblock {\em CoRR}, abs/2112.11927, 2021.

\bibitem{FNBB06}
Eugene Fratkin, Brian~T. Naughton, Douglas Brutlag, and Serafim Batzoglou.
\newblock Motifcut: Regulatory motifs finding with maximum density subgraphs.
\newblock {\em Bioinformatics (Oxford, England)}, 22:e150--7, 08 2006.

\bibitem{GGT89}
G.~Gallo, M.~D. Grigoriadis, and R.~E. Tarjan.
\newblock A fast parametric maximum flow algorithm and applications.
\newblock {\em SIAM J. Comput.}, 18(1):30--55, 1989.

\bibitem{GLM19}
Mohsen Ghaffari, Silvio Lattanzi, and Slobodan Mitrovi{\'c}.
\newblock Improved parallel algorithms for density-based network clustering.
\newblock In {\em Proc.~36th International Conference on Machine Learning (ICML)}, volume~97, pages 2201--2210, 2019.

\bibitem{GhaffariLM19}
Mohsen Ghaffari, Silvio Lattanzi, and Slobodan Mitrovic.
\newblock Improved parallel algorithms for density-based network clustering.
\newblock In {\em {ICML}}, volume~97 of {\em Proceedings of Machine Learning Research}, pages 2201--2210. {PMLR}, 2019.

\bibitem{GKT05}
David Gibson, Ravi Kumar, and Andrew Tomkins.
\newblock Discovering large dense subgraphs in massive graphs.
\newblock In {\em Proceedings of the 31st International Conference on Very Large Data Bases (VLDB)}, pages 721--732, 2005.

\bibitem{Goldberg84}
A.~V. Goldberg.
\newblock Finding a maximum density subgraph.
\newblock Technical Report UCB/CSD-84-171, EECS Department, University of California, Berkeley, 1984.

\bibitem{KhullerS09}
Samir Khuller and Barna Saha.
\newblock On finding dense subgraphs.
\newblock In {\em {ICALP} {(1)}}, volume 5555 of {\em Lecture Notes in Computer Science}, pages 597--608. Springer, 2009.

\bibitem{KS09}
Samir Khuller and Barna Saha.
\newblock On finding dense subgraphs.
\newblock In {\em Proc.~36th International Colloquium on Automata, Languages and Programming (ICALP)}, pages 597--608, 2009.

\bibitem{LancianoMFB24}
Tommaso Lanciano, Atsushi Miyauchi, Adriano Fazzone, and Francesco Bonchi.
\newblock A survey on the densest subgraph problem and its variants.
\newblock {\em {ACM} Comput. Surv.}, 56(8):208:1--208:40, 2024.

\bibitem{snapnets}
Jure Leskovec and Andrej Krevl.
\newblock {SNAP Datasets}: {Stanford} large network dataset collection.
\newblock \url{http://snap.stanford.edu/data}, June 2014.

\bibitem{Manurangsi17}
Pasin Manurangsi.
\newblock Almost-polynomial ratio eth-hardness of approximating densest k-subgraph.
\newblock In {\em {STOC}}, pages 954--961. {ACM}, 2017.

\bibitem{McGregorTVV15}
Andrew McGregor, David Tench, Sofya Vorotnikova, and Hoa~T. Vu.
\newblock Densest subgraph in dynamic graph streams.
\newblock In {\em {MFCS} {(2)}}, volume 9235 of {\em Lecture Notes in Computer Science}, pages 472--482. Springer, 2015.

\bibitem{MitzenmacherV22}
Michael Mitzenmacher and Sergei Vassilvitskii.
\newblock Algorithms with predictions.
\newblock {\em Commun. {ACM}}, 65(7):33--35, 2022.

\bibitem{MNS2025}
Benjamin Moseley, Helia Niaparast, and Karan Singh.
\newblock Faster global minimum cut with predictions.
\newblock {\em CoRR}, abs/2503.05004, 2025.

\bibitem{PQ82}
Jean-Claude Picard and Maurice Queyranne.
\newblock A network flow solution to some nonlinear 0-1 programming problems, with applications to graph theory.
\newblock {\em Networks}, 12(2):141--159, 1982.

\bibitem{karateclub}
Benedek Rozemberczki, Oliver Kiss, and Rik Sarkar.
\newblock {Karate Club: An API Oriented Open-source Python Framework for Unsupervised Learning on Graphs}.
\newblock In {\em Proceedings of the 29th ACM International Conference on Information and Knowledge Management (CIKM '20)}, page 3125–3132. ACM, 2020.

\bibitem{SHKRZ10}
Barna Saha, Allison Hoch, Samir Khuller, Louiqa Raschid, and Xiao-Ning Zhang.
\newblock Dense subgraphs with restrictions and applications to gene annotation graphs.
\newblock In {\em Research in Computational Molecular Biology}, pages 456--472, 2010.

\bibitem{SarmaLNT12}
Atish~Das Sarma, Ashwin Lall, Danupon Nanongkai, and Amitabh Trehan.
\newblock Dense subgraphs on dynamic networks.
\newblock In {\em {DISC}}, volume 7611 of {\em Lecture Notes in Computer Science}, pages 151--165. Springer, 2012.

\bibitem{SJ20}
Saurabh Sawlani and Junxing Wang.
\newblock Near-optimal fully dynamic densest subgraph.
\newblock In {\em Proc.~52th {ACM} Symposium on Theory of Computing ({STOC})}, 2020.
\newblock To appear.

\bibitem{SuV20}
Hsin{-}Hao Su and Hoa~T. Vu.
\newblock Distributed dense subgraph detection and low outdegree orientation.
\newblock In {\em {DISC}}, volume 179 of {\em LIPIcs}, pages 15:1--15:18. Schloss Dagstuhl - Leibniz-Zentrum f{\"{u}}r Informatik, 2020.

\end{thebibliography}

 \end{document}